# Joint Optimization of STAR-RIS Assisted SWIPT Communication Systems

Authors: Junlong Yang，Xizhong Qin，Zhenhong Jia，Jiani Yu
(School of Computer Science and Technology, XJU )

*Abstract*—Simultaneous wireless information and power transfer (SWIPT) is an effective energy-saving technology, but its efficiency is hindered by environmental factors. The introduction of reconfigurable intelligent surfaces (RIS) has alleviated this issue, although it still faces significant constraints due to geographical limitations. This paper proposes a scheme that employs a simultaneously transmitting and reflecting (STAR)-RIS to assist SWIPT. It can overcome this limitation, achieve higher degrees of freedom (DoF), and provide better quality of service (QoS) for users on both sides of the ground. Meanwhile, we have considered a hybrid device based on a power splitting, which is capable of both energy harvesting and information decoding. We have proposed an efficient alternating optimization (AO) method to optimize the phase and amplitude vectors for reflection and transmission, beamforming and the optimal power splitting ratio, achieving an optimal balance between data rate and energy efficiency. Finally, simulation results demonstrate that the sum rate of the proposed model is superior to traditional RIS and other benchmark schemes.

*Keywords-Simultaneous Wireless Information and Power Transfer;Simultaneously Transmitting and Reflecting Reconfigurable Intelligent Surface; Energy Harvesting*

## I. INTRODUCTION

With the progression of 6G standardization, smart cities are experiencing rapid expansion, with a substantial deployment of Internet of Things (IoT) devices across a diverse range of novel applications. The ongoing surge in the quantity of IoT devices, while enhancing the convenience of everyday life, has elevated the requirements for both communication quality and uninterrupted energy provision. The escalating energy consumption is driving researchers towards innovative solutions. Fortunately, SWIPT serves as a potent technology to tackle this challenge and has garnered widespread research attention. SWIPT achieves the dual utilization of radio frequency (RF) signals allowing for the synchronized reception of both data and energy by capturing and absorbing RF energy [1], [2]. This not only guarantees a consistent and reliable energy supply [3] but also enhances the ease of use in wireless networks. The two practical design schemes are time switching and power splitting, where the power splitting (PS) scheme attains an optimal balance between data rate and energy efficiency [4]. Following this, we consider a hybrid device employing the PS protocol to divide the received RF signal into Information Decoding (ID) segments and Energy Harvesting (EH) segments.

However, channel fading represents a notable challenge for SWIPT [5]. Additionally, the presence of obstacles means that Line-of-Sight (LoS) connections may not always be feasible between transmitting and receiving devices. To ensure users receive extensive, high-quality signal coverage, the technology of reconfigurable intelligent surfaces (RIS) emerged [6]. This technology is regarded as a groundbreaking advancement towards creating intelligent wireless environments. [7] investigates the weighted sum rate maximization of an SWIPT system assisted by RIS, while [8] demonstrates that in the same quality of service (QoS) scenario, the BS transmit power is reduced in an SWIPT system assisted by RIS.

While conventional RIS systems have demonstrated effectiveness, studies indicate their limitation in angular range for reflecting incident signals [9]. This necessitates both the BS and users to be positioned on either side of the RIS, allowing coverage for only half of the space. In practice, this geographical constraint poses challenges, significantly limiting the adaptability and effectiveness of RIS deployment. A new concept called simultaneously transmitting and reflecting reconfigurable Intelligent Surface (STAR-RIS) has recently been proposed as a means of overcoming the half space coverage limitation of RIS [10]. STAR-RIS provides simultaneous reflection and transmission of signals from one side, enabling propagation to users on both sides. Every element of the STAR-RIS can be rotated 360°, achieving full-dimensional communication with users regardless of their position relative to the STAR-RIS [11]. Thanks to recently developed prototypes like STAR-RIS [12], [13], highly flexible full-space coverage has become achievable. Harnessing the benefits of full-space coverage, the STAR-RIS-assisted SWIPT system will transcend the geographical constraints of access devices, offering enhanced coverage rates and superior quality support for SWIPT services.

In this paper, we propose a STAR-RIS assisted SWIPT downlink system with the aim of maximizing the downlink sum rate. Specifically, SWIPT employs PS protocol to achieve higher performance. Additionally, the STAR-RIS utilizes the energy splitting protocol, considering not only the phase of reflection and transmission vectors but also their amplitudes. This further enhances system performance and DoF. The intertwining of multiple variables results in a highly coupled and non-convex problem, which makes it challenging to solve.

The contributions of this paper can be summarized as follows:

1) Through the joint optimization of the BS beamforming and the phase and amplitude coefficients of reflection and transmission at the STAR-RIS, the downlink RF energy received by users is maximized.

2) We employ an AO framework, transforming the objective function, and linearizing the non-convex constraints using Lemma 1 [14]. Then, we efficiently solve the subproblems, obtaining suboptimal solutions, and derive the PS ratio at the user, subject to the maximum transmit power constraint at the BS.

3) We implemented the proposed approach and compared it with existing models. The numerical simulation results demonstrate that this model exhibits an overall improvement of approximately 9% compared to the conventional RIS. Moreover, when compared to the system without RIS assistance, the overall performance has improved by 38%.

## II. SYSTEM MODEL AND PROBLEM FORMULATION

### A. System Model

As shown in Fig. 1, we consider a downlink STAR-RIS assisted SWIPT communication scenario, where $N$ elements of STAR-RIS are employed and BS equipped with $M$ single-antennas. There are $K_r$ and $K_t$ users in the reflection and transmission areas of the STAR-RIS, respectively, with $K = K_r + K_t$, $\mathcal{K} = \{1,2...,K\}$. Let $G \in \mathbb{C}^{N \times M}$ denote the channel connecting the BS and STAR-RIS, $\mathbf{f}_i \in \mathbb{C}^{M \times 1}$ denote the beamforming vectors, where $i \in \mathcal{K}$. The user in the reflection area (r) is denoted as $U_{r,k}$, $\mathbf{h}_{r,k}^H \in \mathbb{C}^{1 \times M}$ and $\mathbf{g}_{r,k}^H \in \mathbb{C}^{1 \times N}$ correspond to the channels connecting the BS and STAR-RIS with $U_{r,k}$, where $k \in \mathcal{R} = \{1,2...,K_r\} \in \mathcal{K}$. The user in the transmission area (t) is denoted as $U_{t,k}$, $\mathbf{h}_{t,k}^H \in \mathbb{C}^{1 \times M}$ and $\mathbf{g}_{t,k}^H \in \mathbb{C}^{1 \times N}$ denote channels from the BS and STAR-RIS to $U_{t,k}$, where $k \in \mathcal{T} = \{K_r+1,...,K\} \in \mathcal{K}$. For all channels, we assume the availability of perfect channel state information (CSI) [15].

At the user end, the received signal is divided into two components base on PS ratio: one for ID and the other for EH, where $\rho_{d,k} \in (0,1)$ represent the sections used for ID, and $1-\rho_{d,k}$ denote the remaining portions utilized for EH, $d \in \{r,t\}$. According to the energy splitting protocol, each element of STAR-RIS operates in both transmission and reflection modes, with the ability to optimize individual coefficients. Let $\{\beta_n^r, \beta_n^t \in (0,1)\}$ and $\{\theta_n^r, \theta_n^t \in (0,2\pi]\}$ denote the amplitude and phase shift coefficients of transmission and reflection, respectively, where $n \in \{1,2,\cdots,N\}$. Additionally, the total energy of the transmitted and reflected signals equals that of the incident signal, the constraint $\beta_n^r + \beta_n^t = 1$ must be satisfied. The coefficients vectors for reflection and transmission of the STAR-RIS can be defined as

$$\boldsymbol{\phi}_d = [\sqrt{\beta_1^d}e^{j\theta_1^d}, \sqrt{\beta_2^d}e^{j\theta_2^d}, \cdots, \sqrt{\beta_N^d}e^{j\theta_N^d}]^H. \quad (10)$$

And $\boldsymbol{\Phi}_d = diag(\boldsymbol{\phi}_d)$ represent diagonal coefficients matrices. Thus, the ID section of the received signal at user $U_{d,k}$ is given by

$$y_{d,k} = \sqrt{\rho_{d,k}} \cdot \left[(\mathbf{h}_{d,k}^H + \mathbf{g}_{d,k}^H \boldsymbol{\Phi}_d \mathbf{G}) \cdot \sum_{i \in \mathcal{K}} \mathbf{f}_i x_i + n_{d,k}\right] + z_{d,k} \quad (10)$$

Where $n_{d,k} \sim \mathcal{CN}(0, \sigma_{d,k}^2)$, following a circularly symmetric complex Gaussian distribution with variance $\sigma_{d,k}^2$ and $\gamma_{d,k} = \sigma_{d,k}^{-2}$. $z_{d,k} \sim \mathcal{CN}(0, \delta_{d,k}^2)$ are the additional noise with variance $\delta_{d,k}^2$ produced by the ID section of the signal processing circuit. Therefore, the signal-to-interference-plus-noise ratio (SINR) of received signal at user $U_{d,k}$ can be formulated as

$$SINR_{d,k} = \frac{\gamma_{d,k} \cdot \left|(\mathbf{h}_{d,k}^H + \mathbf{g}_{d,k}^H \boldsymbol{\Phi}_d \mathbf{G}) \cdot \mathbf{f}_k\right|^2}{\gamma_{d,k} \cdot \left[\sum_{j \in \mathcal{K}, j \neq k} \left|(\mathbf{h}_{d,k}^H + \mathbf{g}_{d,k}^H \boldsymbol{\Phi}_d \mathbf{G}) \cdot \mathbf{f}_j\right| + 1\right] + \frac{\delta_{d,k}^2 \gamma_{d,k}}{\rho_{d,k}}}, \quad (3)$$

and the rate of user $U_{d,k}$ is given by $R_{d,k} = \log_2(1 + SINR_{d,k})$.

### B. Problem Formulation

We aim to maximize the sum rate when both total power and minimum EH threshold are constrained. However, practical EH circuits will have losses, which leads us to propose an EH model $E_{d,k}$ that incorporates the realistic conversion efficiency $\eta_{d,k}$

$$E_{d,k} = \eta_{d,k}(1-\rho_{d,k})\left[\sum_{j \in \mathcal{K}} \left|(\mathbf{h}_{d,k}^H + \mathbf{g}_{d,k}^H \boldsymbol{\Phi}_d \mathbf{G}) \cdot \mathbf{f}_j\right|^2 + \sigma_{d,k}^2\right]. \quad (4)$$

Accordingly, the optimization problem is formulated as

$$(P1): \max_{\boldsymbol{\Phi}_d, \mathbf{f}_k, \rho_{d,k}} \sum_{k \in \mathcal{R}} R_{r,k} + \sum_{k \in \mathcal{T}} R_{t,k},$$

$$s.t. \ C1: \sum_{j \in \mathcal{K}} |\mathbf{f}_j|^2 \leq P_{\max},$$

$$C2: E_{d,k} \geq E_{\min},$$

$$C3: \beta_n^t + \beta_n^r = 1, \beta_n^d \in (0,1),$$

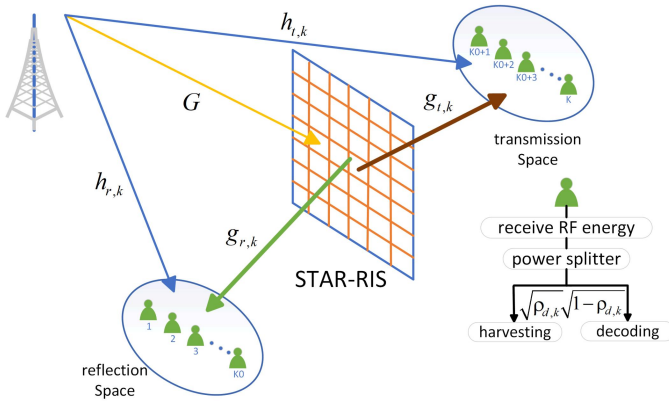

Figure 1. STAR-RIS-aided SWIPT downlink system.

$$C4: \rho_{d,k} \in (0,1), \theta_n^d \in (0, 2\pi], \quad (5)$$

where $P_{max}$ represents the total power constraint of BS and $E_{min}$ represents the minimum EH threshold of users.

## III. ALTERNATING OPTIMIZATION

### A. Optimizing beamforming

For given $\Phi_d$ and $\rho_{d,k}$, we denote $\mathbf{H}_{d,k} = (\mathbf{h}_{d,k}^H + \mathbf{g}_{d,k}^H \Phi_d \mathbf{G})^H (\mathbf{h}_{d,k}^H + \mathbf{g}_{d,k}^H \Phi_d \mathbf{G})$. Define a matrix as $\mathbf{F}_k = \mathbf{f}_k \mathbf{f}_k^H$, then it follows that $rank(\mathbf{F}_k) = 1$ and $\mathbf{F}_k \succeq 0$. To relax the rank-1 constraint, we use the semidefinite relaxation (SDR). Thus, $R_{d,k}$ can be rephrased as

$$R_{d,k} = \log_2 \left( \frac{\gamma_{d,k} \sum_{j \in \mathcal{K}} Tr(\mathbf{H}_{d,k} \mathbf{F}_j) + \frac{\delta_{d,k}^2 \gamma_{d,k}}{\rho_{d,k}} + 1}{\gamma_{d,k} \sum_{j \in \mathcal{K}, j \neq k} Tr(\mathbf{H}_{d,k} \mathbf{F}_j) + \frac{\delta_{d,k}^2 \gamma_{d,k}}{\rho_{d,k}} + 1} \right). \quad (6)$$

By using the logarithmic difference identity and the change of base of logarithms, (6) is rewritten as

$$R_{d,k} \ln 2 = \ln \left( \gamma_{d,k} \sum_{j \in \mathcal{K}} Tr(\mathbf{H}_{d,k} \mathbf{F}_j) + \frac{\delta_{d,k}^2 \gamma_{d,k}}{\rho_{d,k}} + 1 \right) -$$
$$\ln \left( \gamma_{d,k} \sum_{j \in \mathcal{K}, j \neq k} Tr(\mathbf{H}_{d,k} \mathbf{F}_j) + \frac{\delta_{d,k}^2 \gamma_{d,k}}{\rho_{d,k}} + 1 \right). \quad (7)$$

Solving (P1) remains challenging because the objective function is not jointly concave. To address these difficulties, we employ the lemma 1, which offers a potential solution.

**Lemma 1:** Consider the function $\varphi(t) = -tx + \ln t + 1$ for any $x > 0$, then we have

$$-\ln x = \max_{t > 0} \varphi(t) \quad (8)$$

and the optimal solution is $t = 1/x$. Lemma provides an upper bound for $\varphi(t)$, and this bound is tight when $t = 1/x$. Using Lemma 1 and we setting $t_d = S_{d,k}$, $x_d = \gamma_{d,k} \sum_{j \in K, j \neq k} Tr(\mathbf{H}_{d,k} \mathbf{F}_j)$

$+ 1 + \frac{\delta_{d,k}^2 \gamma_{d,k}}{\rho_{d,k}}$, we have

$$R_{d,k} \ln 2 = \max_{S_{d,k} > 0} \varphi_{d,k}(\mathbf{F}, S_{d,k}), \quad (9)$$

where

$$\varphi_{d,k}(\mathbf{F}, S_{d,k}) = \ln \left( \gamma_{d,k} \sum_{j \in \mathcal{K}} Tr(\mathbf{H}_{d,k} \mathbf{F}_j) + \frac{\delta_{d,k}^2 \gamma_{d,k}}{\rho_{d,k}} + 1 \right) -$$
$$S_{d,k} \left( \gamma_{d,k} \sum_{j \in \mathcal{K}, j \neq k} Tr(\mathbf{H}_{d,k} \mathbf{F}_j) + 1 + \frac{\delta_{d,k}^2 \gamma_{d,k}}{\rho_{d,k}} \right) + \ln S_{d,k} + 1. \quad (10)$$

Therefore, the optimization problem can be rewritten as

$$(P2): \max_{S_{r,k}, S_{t,k}, \mathbf{F}} \sum_{k \in \mathcal{R}} \varphi_{r,k}(\mathbf{F}, S_{r,k}) + \sum_{k \in \mathcal{T}} \varphi_{t,k}(\mathbf{F}, S_{t,k}),$$

$$s.t.\ C\ \eta_{d,k}\left(1 - \rho_{d,k}\right)\left[\sum_{j \in \mathcal{K}} Tr(\mathbf{H}_{d,k} \mathbf{F}_j) + \sigma_{d,k}^2\right] \geq E_{min},$$

$$C2: \sum_{k \in \mathcal{K}} Tr(\mathbf{F}_k) \leq P_{max},$$

$$C3: \mathbf{F}_k \succeq 0, \forall k. \quad (11)$$

It can be shown that (P2) is convex. The problem can therefore be efficiently solved using a convex optimization solver, such as CVX. According to Lemma 1, the optimal $(S_{r,k}, S_{t,k})$ can be expressed in closed-form as follows

$$S_{d,k}^* = \left( \gamma_{d,k} \sum_{j \in \mathcal{K}, j \neq k} Tr(\mathbf{H}_{d,k} \mathbf{F}_j) + \frac{\delta_{d,k}^2 \gamma_{d,k}}{\rho_{d,k}} + 1 \right)^{-1}. \quad (12)$$

In (P2), we have relaxed the rank-1 constraint of matrix $\mathbf{F}_k$, but we cannot guarantee its rank-1 property. In case $\mathbf{F}_k$ is indeed rank-1, performing eigenvalue decomposition on $\mathbf{F}_k = \mathbf{f}_k \mathbf{f}_k^H$ would suffice to recover the optimal $\mathbf{f}_k^*$.

Proposition 1: Assuming that all channel links are statistically independent, then $\mathbf{F}_k^*$ satisfies $rank(\mathbf{F}_k^*) = 1$, you can find its proof in [4].

### B. Optimizing $\Phi_t, \Phi_r$

For simplicity, let $\vartheta_d = diag(\mathbf{g}_{d,k}^H) \mathbf{G}$, $\mathbf{v}_d = \phi_d$. Then let $\overline{\mathbf{V}}_d = \left[1, \mathbf{v}_d^H\right]^H$, $\mathbf{V}_d = \overline{\mathbf{V}}_d \overline{\mathbf{V}}_d^H$, it follows that $\mathbf{V}_d \succeq 0$ and $rank(\mathbf{V}_d) = 1$, we also utilize the SDR to handle the rank-1 constraint. Upon denote

$$\mathbf{D}_{r,k} = \begin{pmatrix} \mathbf{h}_{r,k}^H \\ \vartheta_r \end{pmatrix} \sum_{j \in \mathcal{K}} \mathbf{f}_j \mathbf{f}_j^H \begin{pmatrix} \mathbf{h}_{r,k}^H \\ \vartheta_r \end{pmatrix}^H, k \in \mathcal{R}, \quad (13)$$

$$\overline{\mathbf{D}}_{r,k} = \begin{pmatrix} \mathbf{h}_{r,k}^H \\ \vartheta_r \end{pmatrix} \sum_{j \in \mathcal{K}, j \neq k} \mathbf{f}_j \mathbf{f}_j^H \begin{pmatrix} \mathbf{h}_{r,k}^H \\ \vartheta_r \end{pmatrix}^H, k \in \mathcal{R}, \quad (14)$$

$$\mathbf{D}_{t,k} = \begin{pmatrix} \mathbf{h}_{t,k}^H \\ \vartheta_t \end{pmatrix} \sum_{j \in \mathcal{K}} \mathbf{f}_j \mathbf{f}_j^H \begin{pmatrix} \mathbf{h}_{t,k}^H \\ \vartheta_t \end{pmatrix}^H, k \in \mathcal{T}, \quad (15)$$

$$\overline{\mathbf{D}}_{t,k} = \begin{pmatrix} \mathbf{h}_{t,k}^H \\ \vartheta_t \end{pmatrix} \sum_{j \in \mathcal{K}, j \neq k} \mathbf{f}_j \mathbf{f}_j^H \begin{pmatrix} \mathbf{h}_{t,k}^H \\ \vartheta_t \end{pmatrix}^H, k \in \mathcal{T}. \quad (16)$$

Thus, $R_{d,k}$ can be rephrased as

$$R_{d,k} = \log_2 \left( \frac{\gamma_{d,k} Tr(\mathbf{D}_{d,k} \mathbf{V}_d) + 1 + \frac{\delta_{d,k}^2 \gamma_{d,k}}{\rho_{d,k}}}{\gamma_{d,k} Tr(\overline{\mathbf{D}}_{d,k} \mathbf{V}_d) + 1 + \frac{\delta_{d,k}^2 \gamma_{d,k}}{\rho_{d,k}}} \right), k \in \{\mathcal{R}, \mathcal{T}\}$$

(17)

According to Lemma 1: $x = \gamma_{r,k} Tr(\bar{\mathbf{D}}_{r,k}\mathbf{V}_r) + 1 + \frac{\delta_{r,k}^2 \gamma_{r,k}}{\rho_{r,k}}$, $t = q_{r,k}$, $R_{r,k}$ can be expressed as

$$\ln 2 R_{r,k} = \max_{q_{r,k}>0} \psi_{r,k}(q_{r,k}, \mathbf{V}_r) = \ln\left(\gamma_{r,k} Tr(\mathbf{D}_{r,k}\mathbf{V}_r) + 1 + \frac{\delta_{r,k}^2 \gamma_{r,k}}{\rho_{r,k}}\right) - \ln\left(\gamma_{r,k} Tr(\bar{\mathbf{D}}_{r,k}\mathbf{V}_r) + 1 + \frac{\delta_{r,k}^2 \gamma_{r,k}}{\rho_{r,k}}\right), \quad (18)$$

where

$$\psi_{r,k}(q_{r,k}, \mathbf{V}_r) = \ln\left(\gamma_{r,k} Tr(\mathbf{D}_{r,k}\mathbf{V}_r) + 1 + \frac{\delta_{r,k}^2 \gamma_{r,k}}{\rho_{r,k}}\right) - q_{r,k}\left(\gamma_{r,k} Tr(\bar{\mathbf{D}}_{r,k}\mathbf{V}_r) + 1 + \frac{\delta_{r,k}^2 \gamma_{r,k}}{\rho_{r,k}}\right) + \ln q_{r,k} + 1. \quad (19)$$

Similarly, let $x = \gamma_{t,k} Tr(\bar{\mathbf{D}}_{t,k}\mathbf{V}_t) + 1 + \frac{\delta_{t,k}^2 \gamma_{t,k}}{\rho_{t,k}}$, $t = q_{t,k}$, $R_{t,k}$ can be expressed as

$$\ln 2 R_{t,k} = \max_{q_{t,k}>0} \psi_{t,k}(q_{t,k}, \mathbf{V}_t) = \ln\left(\gamma_{t,k} Tr(\mathbf{D}_{t,k}\mathbf{V}_t) + 1 + \frac{\delta_{t,k}^2 \gamma_{t,k}}{\rho_{t,k}}\right) - \ln\left(\gamma_{t,k} Tr(\bar{\mathbf{D}}_{t,k}\mathbf{V}_t) + 1 + \frac{\delta_{t,k}^2 \gamma_{t,k}}{\rho_{t,k}}\right), \quad (20)$$

where

$$\psi_{t,k}(q_{t,k}, \mathbf{V}_t) = \ln\left(\gamma_{t,k} Tr(\mathbf{D}_{t,k}\mathbf{V}_t) + 1 + \frac{\delta_{t,k}^2 \gamma_{t,k}}{\rho_{t,k}}\right) -$$

TABLE I. ALTERNATING OPTIMIZATION FOR SOLVING (P1)

| | |
|---|---|
| Input: | $P_{\max}$, $\sigma_{d,k}$, $\delta_{d,k}$, $L$, $\varepsilon$, $\mathbf{G}, \mathbf{h}_{d,k}, \mathbf{g}_{d,k}$ |
| Output: | $\mathbf{f}_k, \boldsymbol{\phi}_d, \rho_{d,k}$ |
| 1 | Initialize the reflection and transmission coefficients vectors as $\boldsymbol{\phi}_d^{(0)}$, beamforming vector as $\mathbf{f}_k^{(0)}$, power splitter parameter as $\rho_d^{(0)}$. |
| 2 | **Repeat** |
| 3 | Solve problem (11) using the given $\boldsymbol{\phi}_d^{(l-1)}$, $\rho_d^{(l-1)}$ and represent the obtained solution as $\mathbf{f}_k^{(l)}$. |
| 4 | Solve problem (22) using the given $\mathbf{f}_k^{(l)}$, $\rho_d^{(l-1)}$ and represent the obtained solution as $\bar{\mathbf{V}}_d^{(l)}$. |
| 5 | Recovering $\mathbf{v}_d^{(l)}$ from $\bar{\mathbf{V}}_d^{(l)}$ through Gaussian randomization. |
| 6 | Given the values of $\mathbf{v}_d^{(l)}$ and $\mathbf{f}_k^{(l)}$, solve for $\rho_{d,k}^{(l)}$ by (26). |
| 7 | Update $l = l+1$. |
| 8 | **Until** two consecutive iterations are less than the iteration threshold $\varepsilon$ or the number of iterations $l = L$. |

$$q_{t,k}\left(\gamma_{t,k} Tr(\bar{\mathbf{D}}_{t,k}\mathbf{V}_t) + 1 + \frac{\delta_{t,k}^2 \gamma_{t,k}}{\rho_{t,k}}\right) + \ln q_{t,k} + 1. \quad (21)$$

So the problem is reformulated as

$$(P3): \max_{q_{r,k}, q_{t,k}, \mathbf{V}_r, \mathbf{V}_t} \sum_{k \in \mathcal{R}} \psi_{r,k}(q_{r,k}, \mathbf{V}_r) + \sum_{k \in \mathcal{T}} \psi_{t,k}(q_{t,k}, \mathbf{V}_t)$$

$$s.t. \quad C1: \eta_{d,k}(1-\rho_{d,k})\left[Tr(\mathbf{D}_{d,k}\mathbf{V}_r) + \sigma_{d,k}^2\right] \geq E_{\min},$$

$$C2: \mathbf{V}_{t,n,n} + \mathbf{V}_{r,n,n} = 1, \quad n = 2, \cdots, N+1,$$

$$C3: \mathbf{V}_{t,1,1} = \mathbf{V}_{r,1,1} = \frac{1}{2},$$

$$C4: \mathbf{V}_t \succeq 0, \mathbf{V}_t \succeq 0, \quad (22)$$

we can solve it similarly as (P2). After obtaining $\mathbf{V}_r$ and $\mathbf{V}_t$ from $\bar{\mathbf{V}}_r$ and $\bar{\mathbf{V}}_t$ by eigenvalue decomposition with Gaussian randomization, from which we can derive $\mathbf{v}_r = e^{j\arg\left(\left[\frac{\mathbf{V}_r}{\mathbf{V}_{r(N+1)}}\right]_{(1:N)}\right)}$ and $\mathbf{v}_t = e^{j\arg\left(\left[\frac{\mathbf{V}_t}{\mathbf{V}_{t(N+1)}}\right]_{(1:N)}\right)}$.

### C. Derive the optimal PS ratio

When $\boldsymbol{\Phi}_r$, $\boldsymbol{\Phi}_t$ and $\mathbf{f}_i$ are given, the solution for the PS ratio becomes simple since it is decoupled. The optimization of $\rho_{r,k}$ and $\rho_{t,k}$ can be expressed as

$$1 - \rho_{r,k} \geq \frac{E_{\min}}{\eta_{r,k}\left(\sum_{j \in \mathcal{K}}\left|(\mathbf{h}_{r,k}^H + \mathbf{g}_{r,k}^H \boldsymbol{\Phi}_r \mathbf{G})\mathbf{f}_j\right|^2 + \sigma_{r,k}^2\right)}, k \in \mathcal{R}, \quad (23)$$

$$1 - \rho_{t,k} \geq \frac{E_{\min}}{\eta_{t,k}\left(\sum_{j \in \mathcal{K}}\left|(\mathbf{h}_{t,k}^H + \mathbf{g}_{t,k}^H \boldsymbol{\Phi}_t \mathbf{G})\mathbf{f}_j\right|^2 + \sigma_{t,k}^2\right)}, k \in \mathcal{T}, \quad (24)$$

$\rho_{d,k}$ are monotonically increasing according to the objective functions. Clearly, a higher value of $\rho_{d,k}$ is preferable. Based on equation (23) and (24), we can obtain its upper bound.

$$\rho_{d,k}^{(up)} = 1 - \frac{E_{\min}}{\eta_{d,k}\left(\sum_{j \in \mathcal{K}}\left|(\mathbf{h}_{d,k}^H + \mathbf{g}_{d,k}^H \boldsymbol{\Phi}_d \mathbf{G})\mathbf{f}_j\right|^2 + \sigma_{d,k}^2\right)}. \quad (25)$$

Furthermore, based on the defined range of $\rho_{d,k}$, which is $0 < \rho_{d,k} < 1$, we can thereby ascertain the optimal values

$$\rho_{d,k}^* = \min\left\{1, \max\left\{0, 1 - \frac{E_{\min}}{\eta_{d,k}\left(\sum_{j \in \mathcal{K}}\left|(\mathbf{h}_{d,k}^H + \mathbf{g}_{d,k}^H \boldsymbol{\Phi}_d \mathbf{G})\mathbf{f}_j\right|^2 + \sigma_{d,k}^2\right)}\right\}\right\}. \quad (26)$$

### D. Complexity Analysis

The complexity of this algorithm is primarily determined by P2 and P3, L1 and L2 represent the number of iterations for solving the two problems, respectively. Therefore, the algorithm complexity for P2 is denoted as $O\left(L_1 \max\{K,M\}^4 M^{1/2}\right)$, and for P3 it is denoted as $O\left(L_2 \max\{K,N\}^4 N^{1/2}\right)$. Additionally, L3 represents the total number of iterations for Algorithm 2, resulting in an overall algorithm complexity of $O(L_3(L_1 \max\{K,M\}^4 M^{1/2} + L_2 \max\{K,N\}^4 N^{1/2}))$.

## IV. NUMERICAL SIMULATION

In this section, we conduct numerical simulations to evaluate the effectiveness of the model we have put forward. Both the BS and the STAR-RIS are located at (0, 0, 2) meters and (0, 15, 2) meters, respectively. The users on both sides are randomly positioned within circular regions centered at (-2, 15, 1) meters and (2, 15, 1) meters respectively, with a radius of one meter. We set K=4, $K_r = 2$, $\sigma_r^2 = \sigma_t^2 = -70dBm$, $\delta_r^2 = \delta_t^2 = -60dBm$ [16], $P_{max} = 42dBm$. Then, we consider the path loss model as follows $L(d) = \sqrt{C_0 \left(\dfrac{d}{d_0}\right)^{-\alpha}}$, where $C_0 = -30dB$ denotes the reference path loss at $d_0 = 1m$. The path loss exponents of link between BS to STAR-RIS, STAR-RIS to users and BS to users are represented as 2.2, 2 and 3.8 [17], respectively. We assume a Rician channel distribution with Rician factor 3dB for all STAR-RIS assisted links and all direct links follow Rayleigh fading.

Fig. 2 investigates the impact of the number of RIS elements on the sum rate. From Fig. 2, we observe that the "ES mode"

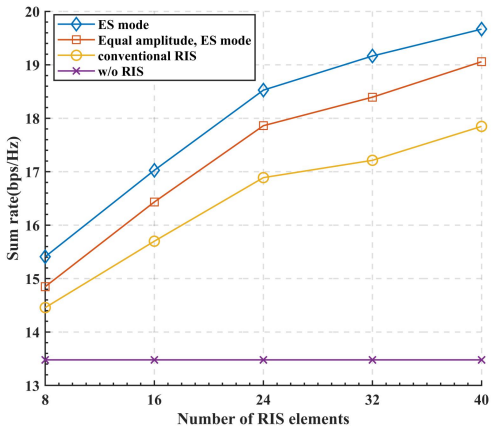

Figure 2. The sum rate varies with the change of N.

scheme, the "Equal amplitude, ES mode" scheme, and the "conventional RIS" scheme show a significant increase with an increase in the number of RIS elements. In contrast, the "without RIS" scheme remains unchanged regardless of the number of RIS elements. It appears that conventional RIS can improve the sum rate of SWIPT systems, and the STAR-RIS with the ES protocol outperforms the conventional RIS in terms of performance. When N = 40, the "ES mode" outperforms the "Equal amplitude, ES mode" by 3.2%, the conventional RIS scheme by 10.2%, and the without RIS scheme by 45.9%.

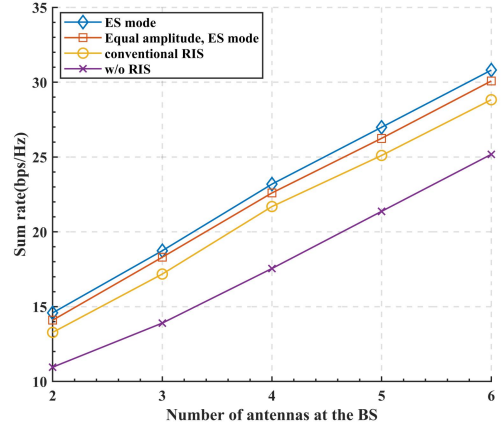

Figure 3. The sum rate varies with the change of M.

In Fig. 3, the number of antennas has an effect on the sum rate. There is a nearly linear growth in the sum rates for all four schemes with an increase in the number of antennas. Furthermore, the proposed "ES mode" scheme consistently outperforms the "conventional RIS" scheme in terms of performance.

As shown in Fig. 4, we investigate the effect of the EH threshold on the sum rate. The sum rate decreases with a stricter EH threshold, and the STAR-RIS provides a greater improvement than conventional RIS. Further, with increasing EH thresholds, the "Equal amplitude, ES mode" ES mode scheme approaches the performance of "ES mode". At this time, it is only necessary to divide the PS ratio equally, and the rate can be close to the maximum rate, thereby reducing the complexity of the algorithm.

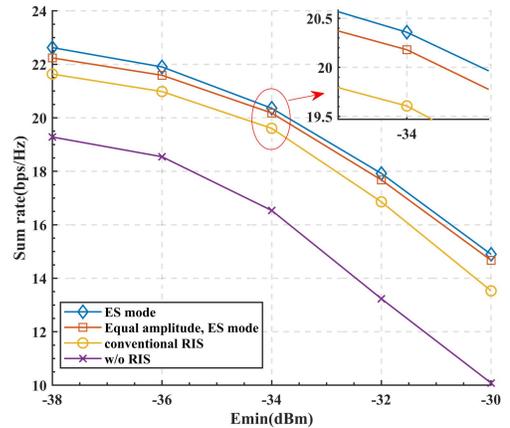

Figure 4. The sum rate varies with the minimum EH threshold.

## V. Conclusion

This letter investigates the maximum achievable sum rate of a SWIPT system assisted by STAR-RIS based on the ES protocol. Furthermore, we provide the corresponding algorithm for this model. A lossy EH model is proposed, combining Lemma 1 jointly optimizes the reflection and transmission coefficients, beamforming coefficients, and PS coefficients. The numerical simulation results demonstrate the performance advantages of this scheme, showing an overall improvement of approximately 9% compared to the conventional RIS, and compared to the system without RIS assistance, the overall performance has improved by 38%.